\documentclass[%
 aip,
pop,%
 amsmath,amssymb,
 reprint,%
]{revtex4-1}

\usepackage{graphicx}
\usepackage{dcolumn}
\usepackage{bm}

\begin{document}

\preprint{AIP/123-QED}

\title[Enhanced Attenuation Arising from Lattice Resonances in a Plasma Photonic Crystal]{Enhanced Attenuation Arising from Lattice Resonances in a Plasma Photonic Crystal}

\author{F. Righetti}
\author{B. Wang}
\author{M.A. Cappelli}
\affiliation{%
Department of Mechanical Engineering, Stanford University, Stanford, CA 94305, USA
}

\date{\today}
\begin{abstract}
We describe the experimental verification of lattice resonances in two-dimensional photonic crystals constructed from an array of gaseous plasma columns.  Enhancements are seen in the extinction of normal incidence transverse electric electromagnetic waves when the localized surface plasmon modes of the plasma columns are shifted into the vicinity of the photonic crystal Bragg resonances. Simulations and experiments are in reasonable agreement and confirm the appearance of a Fano-like profile with deep and broad extinction bands. The broadening of the spectra as surface plasmon modes come into coincidence with Bragg gaps suggest that the Bragg fields couple strongly into the radiating Mie dipoles to drive enhanced damping of the photonic crystal resonance.

\end{abstract}

\pacs{Valid PACS appear here}
\maketitle

\section{Introduction}

The properties of plasma photonic crystals have attracted considerable attention
because they afford an element of control that is generally not available with more conventional metallic or dielectric photonic crystal variants \cite{hojo2004dispersion,sakai2005verification,fan2010tunable, wang2016plasma, wang2016waveguiding}. In photonic crystals, the geometry and electromagnetic properties of the structures define the position, the width, and the depth of the band gaps.\cite{joannopoulos2011photonic} A rapidly modifiable bandgap would be a desirable feature and although methods of reconfiguring conventional photonic crystals have been demonstrated \cite{kitagawa2012thz,chong2004tuning,yang2011nanoscale}, those involving either mechanical or thermal manipulation of the structure have a relatively low response frequency. A plasma photonic crystal, with a variable refractive index, affords the possibility of rapidly controlling the bandgap at rates limited only by the time to form or recombine the plasma. 

In a previous study we described the behavior of relatively narrow and deep bandgaps arising from the excitation of localized surface plasmons (LSP) in two dimensional photonic crystals comprised of gas-discharge driven plasma columns \cite{wang2016plasma}. The depth of these LSP bandgaps was found to be much greater than the photonic (Bragg) bandgap because of the limited size of the crystal (7 x 7 square array). The frequency of these LSP gaps was varied by changing the plasma density through control of the discharge current. In a subsequent study \cite{wang2016waveguiding} we demonstrated electromagnetic (EM) waveguiding along line vacancies using these LSP bandgaps, even within a plasma photonic crystal of limited size. We showed that the plasma array and waveguide geometry is easily reconfigured by independently powering the plasma columns demonstrating the potential for the rapid steering of incoming electromagnetic waves. 
\begin{figure}
\includegraphics[width=9cm,height=9cm,keepaspectratio]{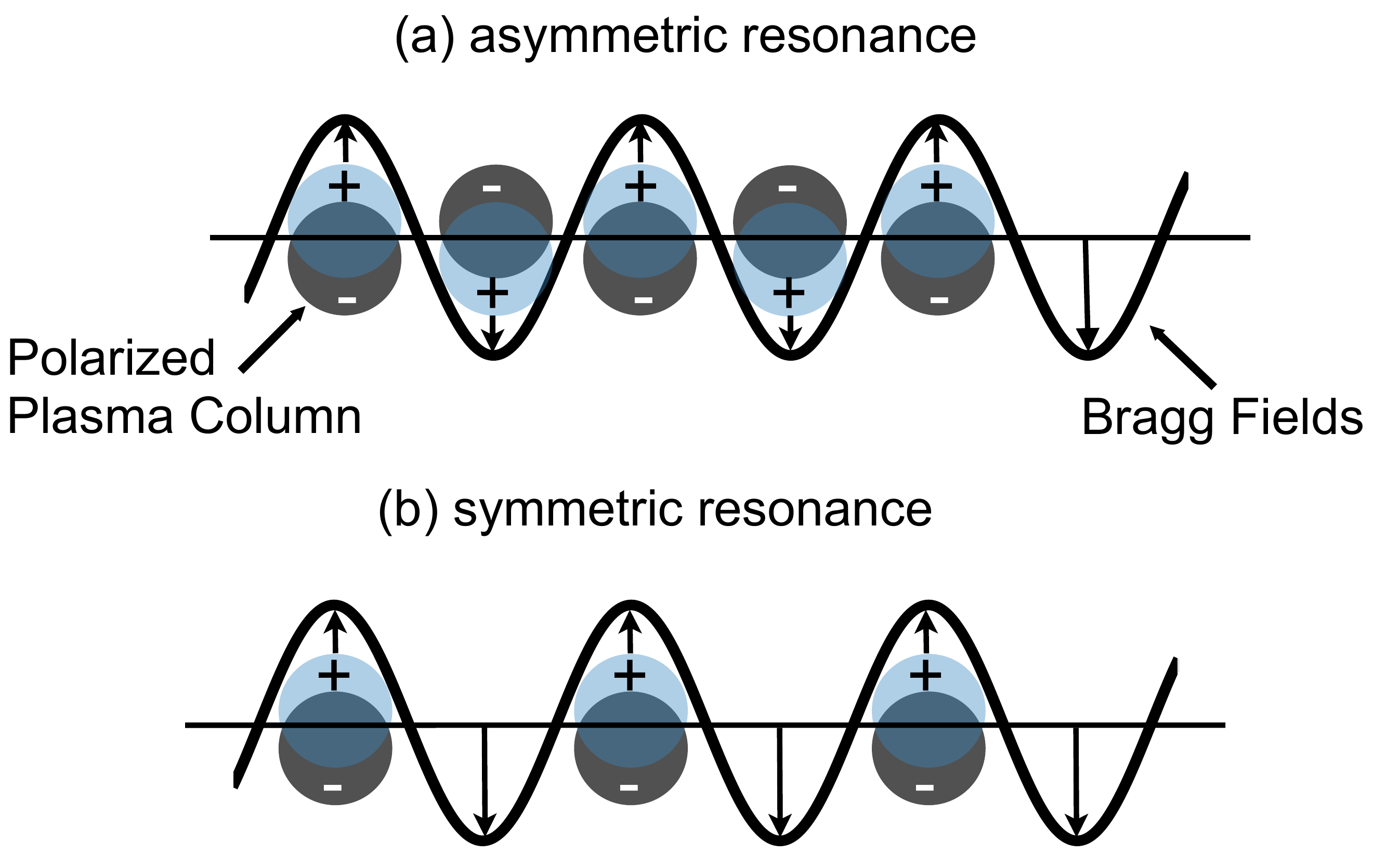}
\caption{\label{fig:braggfield} Schematic illustration of lattice resonance. (a) half-wavelength resonance with adjacent plasma columns asymmetrically driven. (b) full-wavelength resonance with adjacent columns symmetrically driven.}
\end{figure}
%

This letter addresses the exploitation of lattice resonances \citep{humphrey2014plasmonic} to control the depth and width of these bandgaps. Lattice resonances arise from the excitation of localized surface plasmons in periodic structures when the dipole fields of the LSP constructively or destructively interfere with the Bragg scattered fields.  A qualitative depiction of two examples of such resonances is shown in Fig. 1.  When the lattice spacing $a$, coincides with half of the Bragg field wavelength, $\lambda$, the adjacent plasma columns are asymmetrically polarized (Fig.~\ref{fig:braggfield}(a)). Conversely, when $a = \lambda $, the adjacent fields will be symmetrically polarized (Fig. 1(a)). Lattice resonances are commonly seen in metallic photonic crystals \cite{rodriguez2012collective, chu2008experimental}.  They give rise to a Fano extinction/absorption spectrum which is a characterization of the response of a pair of coupled driven linear oscillators with closely spaced resonance frequencies but differing dissipation \cite{klein2005lineshape}. Often referred to broadly as Fano resonances \cite{giannini2011fano, miroshnichenko2010fano}, the theory for these resonances was originally developed to describe quantum mechanical (atomic) systems\cite{fano1961effects}, but has since been applied quite broadly to a range of physical problems and is often explained using classical mass-spring analogues\cite{joe2006classical}.

One feature of lattice resonances in metallic photonic crystals is the presence of resonance peak enhancement and narrowing in the extinction spectrum \cite{chu2008experimental}.  The interactions near resonance between the scattering field in photonic crystals and the dipole fields associated with surface plasmons can result in both enhanced or reduced radiative damping. Enhanced damping when in resonance introduces broadening of the extinction band. This may provide an opportunity to design improved wave-guiding structures of limited scale, as the Bragg gaps in plasma photonic crystals of limited size (i.e., limited number of elements) are rather weak and the surface plasmon gaps, while stronger, are still not sufficiently broad to see practical field steering of incident pulsed fields which have Fourier components that might span a range of frequencies.

We present below, through simulations and experiments, evidence of this lattice (Fano) resonance in a plasma photonic crystal. The ensuing resonance enables opportunities in the ability to rapidly reconfigure the response of the plasma photonic crystal, in particular, the ability to modify its extinction behavior in the vicinity of the Bragg gaps by the relative ease of changing the plasma electron density - a characteristic that is not achievable with the metallic photonic crystal variant. 

We first discuss simulations carried out on a two-dimensional laterally infinite but four-deep square lattice of uniform plasma columns surrounded by vacuum to illustrate the presence of this lattice resonance. The dielectric constant of the plasma columns is described by a Drude model:       
\begin{equation}
\varepsilon_{p} = 1 - \frac{\omega_{p}^{2}}{\omega(\omega+i\nu)}
\end{equation}

Here, $\omega$ is the EM wave frequency, $\nu$ is the collisional damping frequency, and $\omega_p$ is the plasma frequency, given by $\omega_p$ = $\sqrt{n_e {e}^{2}/{m}_e \varepsilon_o}$. For a collisionless plasma ($\nu = 0$), $\varepsilon_p$ takes on values near unity well above $\omega_p$, crosses through zero at $\omega_p$, and then becomes strongly negative just below $\omega_p$. In these simulations, we take the ratio of the plasma column radius to lattice constant, $r/a = 0.4$. Other important frequencies which we can define for this problem include the lattice frequency, $\omega_l = \pi c/a$ (corresponding to a half-wavelength resonance between the crystal elements) and the localized surface plasmon frequency, which for polarization within a plane perpendicular to the column is $\omega_{sp} = $ ${\omega_p}/\sqrt{2}$ (lowest Mie resonance mode). In the presentation of the results, we non-dimensionalize all frequencies by the lattice frequency, $\omega_l$. We vary the plasma frequency (which in practice is achieved by varying the plasma density) holding the ratio $\nu$/$\omega_l$ constant. For these initial simulations we consider a very low collisionality plasma, with $\nu$/$\omega_l = 0.005$.
\begin{figure}
\includegraphics[width=13cm,height=13
cm,keepaspectratio]{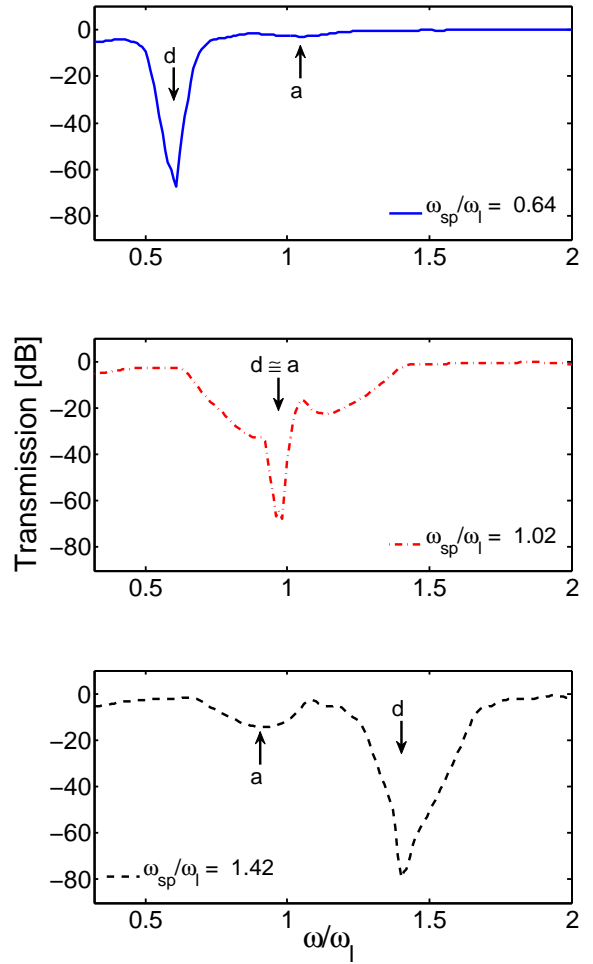}
\caption{\label{fig:epsart} HFSS simulations of TE wave transmission along the $\Gamma - X$ direction of a 4 x infinite array of plasma columns. Top to bottom panels vary the plasma density resulting in a non-dimensionalized surface plasmon frequency of  $\omega_{sp} / \omega_l =$ 0.64 (top), 1.02 (center) and 1.42 (bottom). The identifiers $d$ and $a$ represent the location of the LSP gap and the Bragg bandgap respectively.}
\end{figure}
The plasma photonic crystal is subjected to incident EM waves of transverse electric (TE) polarization so as to excite the LSP resonances within the plasma columns and the crystal is oriented so that the propagation is along the $\Gamma - X$ direction in reciprocal lattice space. We use the ANSYS commercial software package generally referred to as the High Frequency Electromagnetic Field Simulator (formally known as the High Frequency Structural Simulator), or specifically, ANSYS HFSS 16, to compute the transmission spectra. The plasma is simulated as a dielectric rod of a dielectric constant given by Eqn. 1 above. Representative transmission spectra are shown in Fig. 2 for three values of the plasma density, $n_{e}$, resulting in a non-dimensionalized surface plasmon frequency of  $\omega_{sp} / \omega_l =$ 0.64 (Fig. 2 (a),  $\omega_{sp} / \omega_l =$1.02 (Fig. 2(b)) and  $\omega_{sp} / \omega_l =$1.42 (Fig. 2(c)). Apparent from the figures is the fact that the first Bragg gap (designated by the arrow with label ``a") is rather shallow and its location is relatively insensitive to the plasma density. In contrast, the LSP gap (designated by the label ``d") is quite deep, and its location varies linearly with the plasma frequency ($\sim n_{e}^{1/2})$. It is noteworthy that the location of this Bragg gap is below the plasma frequency and so there is strong reflection at the vacuum-plasma interface and the fields within the plasma columns are evanescent. In Fig. 2 (a), the Bragg gap, expected to be at $\omega / \omega_l =$ 1, appears to peak at slightly higher frequency, presumably due to a Fano anti-resonance\cite{miroshnichenko2010fano} at a frequency of $\omega / \omega_l =$0.85.  The lattice resonance is clearly evident when the LSP and Bragg gaps coincide (Fig. 2 (b)). Here, we see the characteristic Fano profile\cite{miroshnichenko2010fano} superimposed on a broadened underlying background. When the plasma density is varied such that the LSP is swept past the Bragg gap to higher frequencies (Fig. 2 (c)), the apparent peak in the Bragg bandgap shifts to $\omega / \omega_l < $1 due to the anti-resonance which now appears at approximately $\omega / \omega_l =$1.1.

Experiments were carried out for a 7 x 7 square lattice plasma photonic crystal a schematic of which is shown in Fig. 3. The crystal is comprised of an array of discharge tubes that can generate plasma densities corresponding to plasma frequencies as high as $\omega_p =6 \times 10^{10}$ rad/s (or non-dimensionalized plasma and LSP frequencies as high as $\omega_p /\omega_l =1.5$, or $\omega_{sp} /\omega_l =1.1$, respectively).  The details of its construction have been described in previous papers \cite{wang2016plasma, wang2016waveguiding}. The variable plasma density of these discharge tubes were studied by using a single discharge tube in the vacancy defect of a 2D photonic crystal comprised of alumina rods \cite{wang2015tunable}.

\begin{figure}
\includegraphics[width=8cm,height=8cm,keepaspectratio]{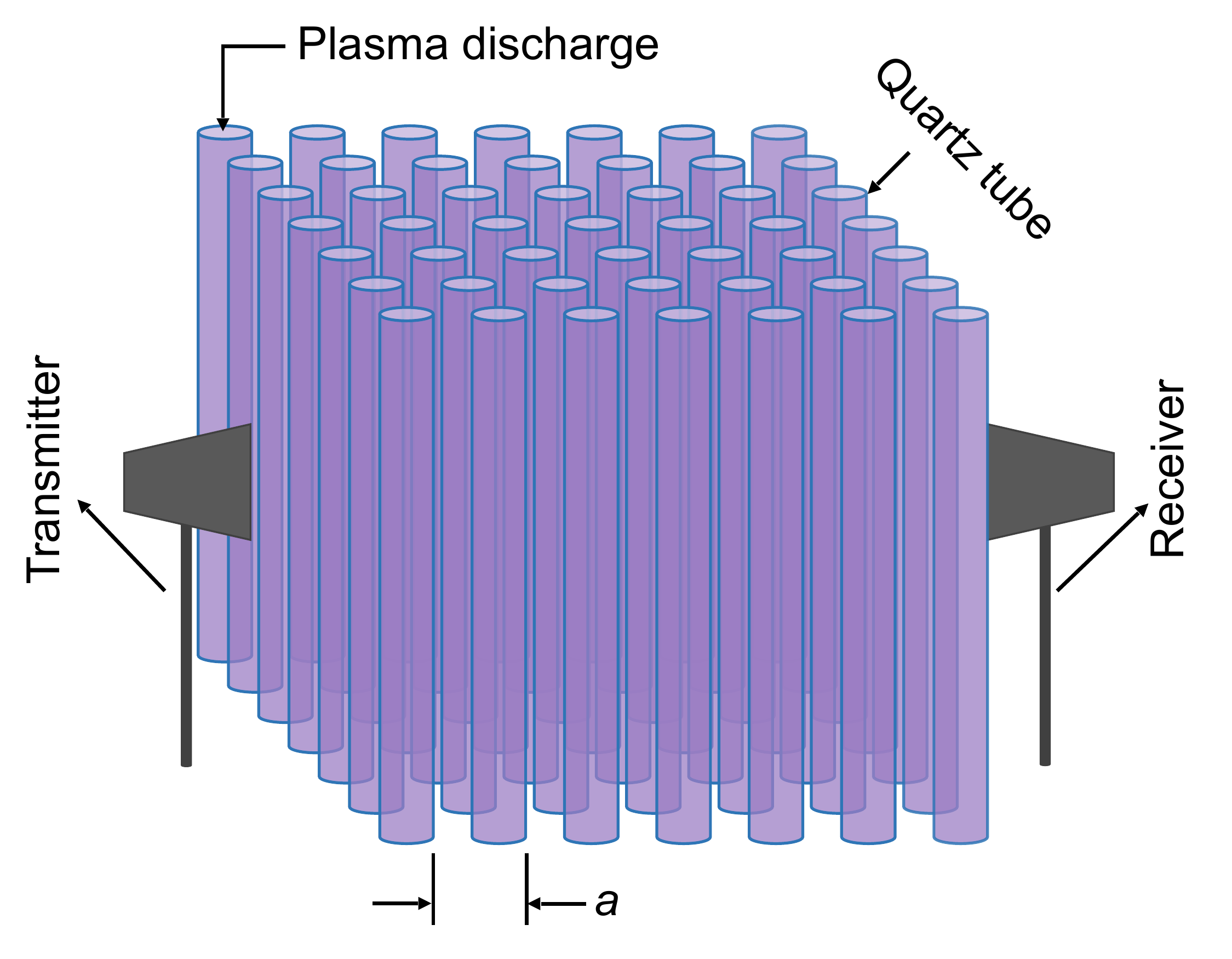}
\caption{\label{fig:epsart} Plasma discharge tubes arranged in a 7 x 7 square lattice with transmitting and receiving microwave horns arranged to study EM wave propagation along the $\Gamma - X$ direction.}
\end{figure}

The plasma columns are sustained by an external discharge excitation of  argon/mercury vapor at a pressure of about 250 Pa within a 15 mm diameter, 290 mm long quartz tube of 1 mm wall thickness. The discharges
are driven by an AC ballast with a triangular wave form output and peak to peak voltage of 160V. The associated ballast had a variable peak current (also close to triangular in wave form) ranging from 25 mA to 111 mA, with a ballast frequency that decreased from 55.0 kHz to 32.0 kHz for increasing peak current. Control of the plasma current by variable control of the input ballast voltage allows us to access electron densities as high as about $n_{e} = 10^{12}$ cm$^{-3}$ . To study transmission through the array we use broadband transmitting and receiving horns (A-INFO Model LB-20180-SF) that can span the S through Ku bands of the EM spectrum (2$-$18 GHz). An HP 8722D Vector Network Analyzer is used to determine the $S_{21}$ parameter which characterizes the transmission coefficient of the device. For HFSS simulations of the transmission through the array, we assume that the density within the plasma columns are uniform over a radius of 4.5 mm. This radius, which is smaller than the radius of the inner volume of the quartz tube, accounts for the non-uniform variation in plasma density expected due to diffusion and loss to the wall and represents the average density of a parabolic radial density profile (of higher centerline density) over a diameter of 13 mm.  Finally, we note that the lattice constant of 25 mm (corresponding to a lattice frequency $\omega_l = 3.8\times 10^{11}$ rad/s or $f_l=\omega_l/2\pi = 6 $ GHz) chosen here is different than that of previous studies, so as to enable the overlapping of the Bragg and LSP gaps topping of drive a lattice resonance. For the simulations, also described below, we therefore use a column radius to lattice constant ratio, $r/a = 0.18$.

Experimental results of the measured transmission of the plasma photonic crystal at conditions in which the plasma is inactive and for two values of estimated plasma densities are shown in Fig. 4 (solid red lines). When the plasma is off (Fig. 4(a)), the recorded spectrum serves as a reference, and contains features labeled as ``a", ``b", and ``c", which are attributed to weak bandgaps associated with the quartz tubes alone. At a relatively low plasma density of $n_{e} = 2\times  10^{11}$ cm$^{-3} $ (Fig.4 (b)), we see the clear emergence of an LSP gap (labeled as ``d" in the figure) at approximately $\omega_{sp} /\omega_l =0.45$. The non-dimensional plasma frequency is located at  $\omega_p /\omega_l =0.69$, just below the first bandgap. At this density, the presence of the plasma seems to have a negligible effect on the bandgaps of what is now a composite quartz-plasma (binary structure) photonic crystal. 
 
\begin{figure}
\includegraphics[width=13cm,height=13cm,keepaspectratio]{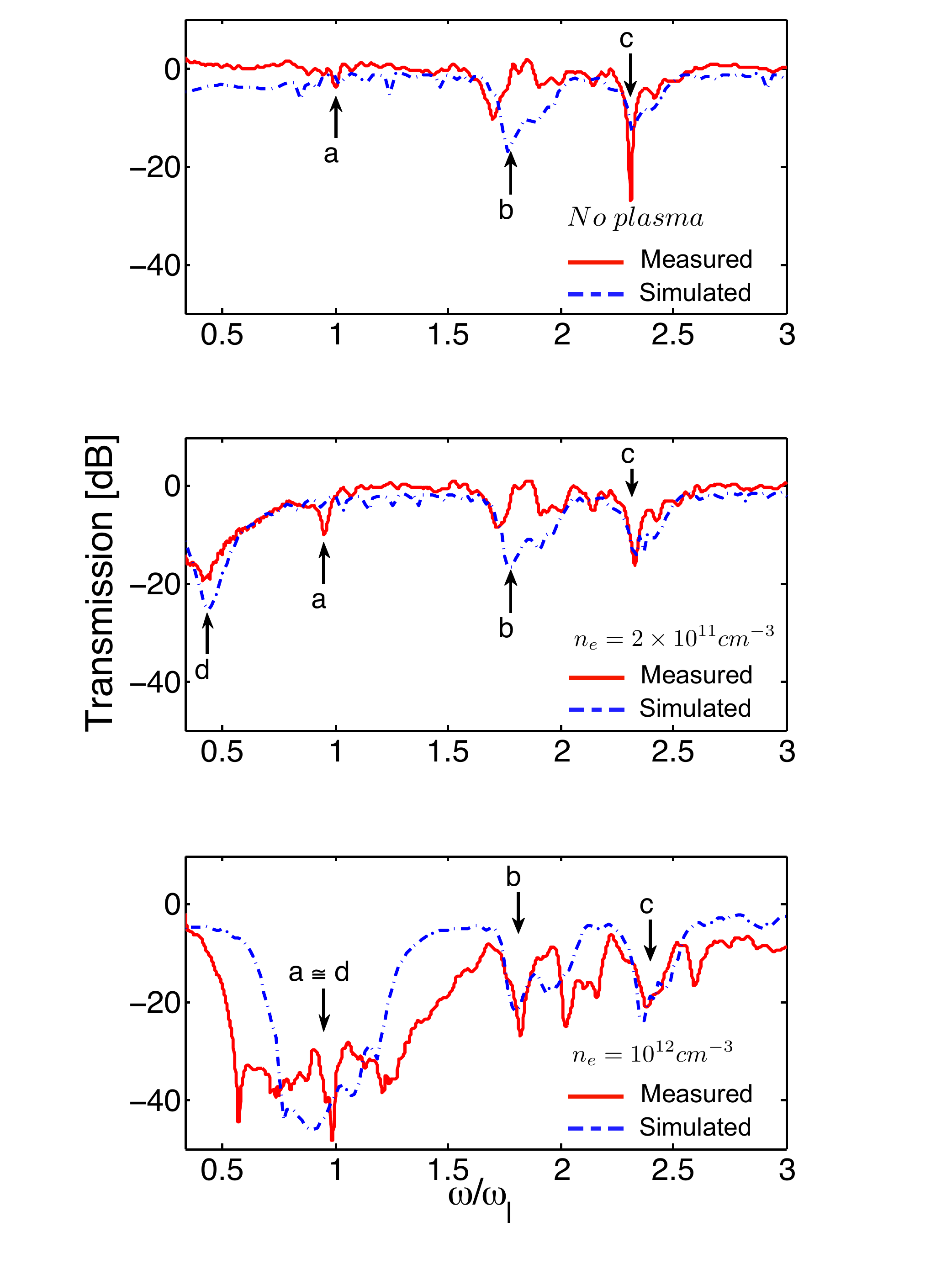}
\caption{\label{fig:epsart} TE wave transmission along the $\Gamma - X$ direction of the 7 x 7 discharge tube array. (a) plasmas are inactive. Labels refer to the location of the first three bandgaps of the quartz tube array. (b) plasma active at low plasma density and the label ``d" points to the location of the LSP gap. (c) high plasma density case where the LSP mode coincides with the first bandgap. In all cases, HFSS simulations using a collision frequency $\nu/\omega_l$ = 0.17.}  
\end{figure}

When the plasma density is elevated to about $n_{e} = 10^{12} $ cm$^{-3}$ where the LSP and Bragg gaps overlap, we see a strong interaction that leads to a deep and broad gap, with underlying structure (Fig. 4(c)).  The broadening of the spectra as the LSP gap comes into coincidence with Bragg gap suggests that the Bragg fields couple strongly into the radiating Mie dipoles to drive enhanced damping of the photonic crystal resonance. In comparison to the lower density case, the depth of the gap increases by a few orders of magnitude. We see evidence of a broadening/splitting of the underlying spectrum as well as a central Fano profile similar to that simulated in Fig. 2(b), although less pronounced. At these densities, the first Bragg gap is below the plasma frequency.  However, the higher photonic crystal bandgaps are now in proximity with the plasma resonance, with $\omega_p /\omega_l =1.5$ and an enhancement is also seen in the depth and width of those gaps. 
Also shown in Fig. 4, as dashed blue lines, are the results of ANSYS HFSS simulations of our finite plasma photonic crystal array, including the quartz envelope surrounding the plasma columns. The simulations require specifying the collisional damping rate, which is not known precisely. We find that $\nu = 1$ GHz ($\nu/\omega_l$ = 0.17) is a reasonable value based on the gas pressure and leads to relatively good agreement with the measured transmission spectra. Simulations carried out at lower values for the collision frequency produced a deeper gap and a more narrow spectrum at the lattice resonance condition. While a higher value led to better agreement with the gap width, it resulted in too high of a transmission on resonance.  

This demonstration of lattice (Fano) resonances in plasma photonic crystals results in greatly-enhanced bandgaps when the plasma photonic crystal is of limited size. We have shown that the width of the bandgap is enhanced substantially, resulting in the possibility of transmission control over a broad range of frequencies.  We also see reasonable agreement between experiments and simulations. The finer structure seen in the experiments during resonance needs to be studied more carefully and is the subject of a future paper. The presence of the quartz tubes lead to relatively shallow bandgaps, but nevertheless mask the bandgaps associated with the plasma itself.  Future experiments aim at generating unconfined plasma arrays so as to better reveal the bandgap structure without the quartz interference.  

This research was supported by a Multidisciplinary University Research Initiative from the Air Force Office of Scientific Research, with Dr. Mitat Birkan as the program manager. The authors would like to thank U. Shumlack and W. Thomas for stimulating discussions.   

\bibliography{aipsamp}

\end{document}